\documentclass[preprint,showpacs,apl,preprintnumbers,superscriptaddress,amsmath,amssymb]{revtex4-1}
\usepackage{graphicx,dcolumn,bm,color}

\begin{document}

\title{Disorder and electron interaction control in low-doped silicon metal-oxide-semiconductor field effect transistors}

\author{T.~Ferrus \footnote{Present address : Hitachi Cambridge Laboratory, J. J. Thomson Avenue, CB3 0HE, Cambridge, United Kingdom}}
\email{taf25@cam.ac.uk}
\author{R.~George}
\author{C.~H.~W.~Barnes}
\author{M.~Pepper}

\affiliation {Cavendish Laboratory, University of Cambridge, J. J. Thomson Avenue, CB3 0HE, Cambridge, United Kingdom}

\date{\today}

\begin{abstract}

We fabricated silicon metal-oxide-semiconductor field effect transistors where an additional sodium-doped layer was incorporated into the oxide to create potential fluctuations at the Si-SiO$_2$ interface. The amplitude of these fluctuations is controlled by both the density of ions in the oxide and their position relative to the Si-SiO$_2$ interface. Owing to the high mobility of the ions at room temperature, it is possible to move them with the application of a suitable electric field. We show that, in this configuration, such a device can be used to control both the disorder and the electron-electron interaction strength at the Si-SiO$_2$ interface.

\end{abstract}


\newpage
\maketitle

The presence of disorder in materials is a key element in understanding various types of electrical or magnetic behavior - especially electronic phase transitions. Disorder is often associated with a non-homogeneous distribution of magnetic or non-magnetic impurities, the existence of dislocations in the crystal or defects due to a lattice mismatch at interfaces. It is responsible for the destruction of the superconducting state in Bechgaard salts,\cite{Choi} and the existence of weak localization and Anderson localization in semiconductors,\cite{Anderson} the metal-insulator transition \cite{Hamilton} or the Mott to Efros regime crossover.\cite{Crossover} Disorder also causes the formation of glass-like electronic states when the pinning energies become larger than the elastic energy of a Wigner crystal.\cite{Wigner,Pepper1} An external magnetic field or pressure could be used to control the effective disorder in an electronic system but in-situ control is more difficult. In metal-oxide-semiconductor field effect transistors (MOSFETs), the use of a top and back gate allows one to control both the electron-electron interaction and the disorder but only the relative strength can be experimentally accessed as the effective disorder felt by the carriers depends on their density. Mott \textit{et} al \cite{Mott3} showed that metal-nitride-oxide-semiconductor (MNOS) devices are useful as the potential fluctuations and the disorder at the Si-SiO$_2$ interface are influenced by the presence of the thin oxide-nitride interface. However, the position of the Si$_3$N$_4$ layer, and so, the disorder strength were determined during the wafer growth. 

In this paper, we describe a device based on MOSFET technology where a layer of sodium impurities was introduced into the silicon oxide barrier. At room temperature, this layer was mobile and could be moved to different positions in the oxide using an electric field, allowing us to control the disorder at the interface in a single device. Additionally, the top metal gate could be used to control the strength of the electron-electron interaction.

The device fabrication follows a standard silicon MOSFET process \cite{Ferrus1} but requires the disorder control layer (DCL) to be introduced prior to the gate metalization. By depositing a solution of sodium chloride onto the surface of the silicon oxide, an Helmholtz layer \cite{Helmholtz} is formed with the Cl$^-$ions lying in the liquid outer layer and the Na$^+$ ions attracted and diffusing in the oxide.\cite{Snow} At low concentration ($\sim 10^{-7}$ mol$^{-1}$), the ion concentration in the oxide is controlled by adjusting the time over which the solution is in contact with the oxide. The high activation energy of the ions in SiO$_2$ and their high diffusivity at room temperature allows the DCL to be easily positioned in the oxide for reliable device operation at low temperature. Lithium or potassium could also be used as alternatives because of their similar properties.
We previously showed that the conductivity of these devices is dominated by hopping conduction typically below 20\,K and that the hopping exponent $s$ has an unusual variation with gate voltage and position of the DCL in the oxide.\cite{Ferrus2} At a specific DCL distance $h$ from the Si-SiO$_2$ interface, the decrease of $s$ with electron density $n_s$ is explained by the presence of long conduction band tails due to the presence of impurity states at the Si-SiO$_2$ interface and a decrease in the localization length with electron density. The observed crossover from Efros-Shklovskii variable range hopping (ES VRH) \cite{Efros1} and Mott VRH \cite{Mott} when the DCL is moved closer to the interface occurs when potential fluctuations whose amplitude is proportional to $1/h^2$ and whose length scale varies as $h$ \cite{Efros2} become strong enough so that short-range disorder and electron screening \cite{Gumbs} prevail, leading to the suppression of the soft Coulomb gap in the density of states.\cite{Shklovskii2}

Our previous results suggest a complex phase diagram in the electron density-disorder plane (Fig. 1). This explains the apparent inconsistency between various experimental results obtained on silicon MOSFETs, either presenting the characteristics of a Mott VRH \cite{Hartstein,Timp,Pepper1} or a ES VRH \cite{Mason-Mertes} or a ES to Mott transition as function of temperature.\cite{Ishida} ES VRH is likely to be present in devices where the potential fluctuations at the interface are small in amplitude and consequently in high-mobility MOSFETs. This is the case for our undoped devices \cite{Ferrus1} or in the present devices when the ions are moved away from the interface. The presence of impurities in the oxide decreases the mobility of MOSFETs drastically, especially when they are close to the interface, leading to the absence of correlated hopping. For completeness, one needs to address the issue of the ion concentration in the oxide. Previous studies on similar but low-sodium-doped MOSFETs have shown a hopping VRH with a $T^{-2/5}$ dependence on the conductivity over a wide range of gate voltages.\cite{Ferrus1} This was attributed to a screened ES VRH caused by the presence of a metal gate. This situation arises when the ions were close to the Si-SiO$_2$ interface, showing that the present phase diagram may change with ion concentration. The importance of the ion concentration and distribution across the oxide is also illustrated by the existence of an impurity band \cite{Timp} or Hubbard bands \cite{Ferrus1} in some devices and by the presence of conductivity fluctuations in others. Preliminary studies seem to indicate the existence of impurity bands only for a restricted range of ion concentration and position of the DCL in the oxide.

\begin{figure}
\begin{center}
\includegraphics[width=85mm,bb=3 1 241 128]{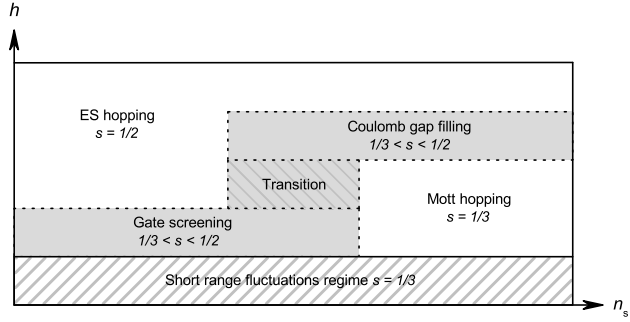}
\end{center}
\caption{\label{fig:figure1} Observed phases in a silicon MOSFET with DCL, as a function $n_s$ and $h$.}
\end{figure}

To assess the ability of these structures to control disorder and electron-electron interaction, it is essential to get an insight into the variation of the disorder strength with the position of the DCL. This can be achieved by estimating the value of the minimum non-activated conductivity $\sigma_{\textup{min}}$ for each drift \cite{Mott2}. It should be noted that, although all states are localized and consequently a minimum metallic conductivity does not strictly exist as conductivity is zero at absolute zero. Nevertheless at finite temperatures, the phase coherence length is of the order or less than the elastic mean free path. Hence, the localization is cut off and the states are effectively extended so that the concept of minimum metallic conductivity remains a useful experimental description. Its value is usually obtained by scaling $\sigma\left(T\right)$ at different $V_{\textup{g}}$ onto a same curve.\cite{Kravchenko} However such a scaling is only appropriate when the device is close to the metal-to-insulator transition, a region that is far beyond the range of electron density investigated in this paper. Deep in the insulating region of 2D systems, literature provides very little information on $\sigma_{\textup{min}}$. Nevertheless, $\sigma\left(T\right)$ is well modeled by

\begin{eqnarray}\label{eqn:equation1}
\sigma\left(T\right) = \sigma_{\textup{min}}\,\left(T_s/T \right)^{2 s}\,\textup{e}^{-\left(T_s/T \right)^{s}}
\end{eqnarray}

Above the threshold voltage, the minimum conductivity saturates at a value close to 0.18 e$^2$/h independently of the ion position (Fig. 2a), except when the DCL is near the metal gate ($h \sim d_{\textup{ox}}$) for which $\sigma_{\textup{min}}\,\sim\,0.37 e^2/h$. Indeed, in this case, $\sigma_{\textup{min}}$ decreases with electron density (Fig. 3a). This gives evidence for the presence of long range potential fluctuations \cite{Pepper2} which invalidates the derivation of $\sigma_{\textup{min}}$ based on the Anderson model. In this case, the Fermi surface in silicon is expected to be circular. By positioning the DCL closer to the Si-SiO$_2$ interface, the amount of short-range disorder is increased, as shown by the increasing potential disorder at low electron density. This, however, may not be true and the potential not completely random at the Si-SiO$_2$ interface ($h \sim 0$). The Fermi surface then becomes distorted due to scattering anisotropy at the interface and the band degeneracy in silicon \cite{Altshuler} is lifted. So we expect $g_v\,=\,2$ when $h \sim d_{\textup{ox}}$ whereas $g_v\,=\,1$ in the other cases. The observed value for $\sigma_{\textup{min}}\,\sim\,0.37 e^2/h$ at high electron density is smaller than already reported values in 2D systems \cite{smin1} but is consistent with Thouless' criterion for localization.\cite{Mott3} Such a low value is often a sign of exponentially localized states at the interface. The presence of charge in the oxide is responsible for the decrease of $\sigma_{\textup{min}}$  \cite{Ando} and it is expected that $\sigma_{\textup{min}}\,\sim\,0.37\, e^2/h$ corresponds to a charge oxide density of $3\,10^{12}$cm$^{-2}$ which is close to the estimate value from the threshold shift.

By definition of the minimum conductivity, we also have

\begin{eqnarray}\label{eqn:equation2}
\sigma_{\textup{min}}= g_{\textup{v}} \frac{e^2}{h} \left(k_{\textup{F}}l \right),
\end{eqnarray}

with $k_{\textup{F}}$ the Fermi wavevector, $l$ the electron mean free path and $g_{\textup{v}}$ the band degeneracy.

The value of $\left( k_{\textup{F}}l \right)^{-1}$ is usually considered as a direct measure of the disorder strength. Above $\Delta V_{\textup{g}}\,\sim\,0.05$ V, it is quite insensitive to the position of the DCL. At lower density, it increases when $h$ decreases in agreement with an increase of disorder when the effective oxide charge increases at the Si-SiO$_2$ interface (Fig. 2b). However, the disorder decreases for $h > 30$ nm. This is a clear indication of the existence of an optimum position of the DCL for which the disorder is minimum. This is in contradiction with Mott's concept of minimum metallic conductivity where $\sigma_{\textup{min}}$ is expected to be relative insensitive to potential fluctuations in the system. When the ions are close to the SiO$_2$-Al interface, electrons are weakly localized away from the interface. Disorder then mostly results from scattering between impurity states. On the contrary, when close to the Si-SiO$_2$ interface, electrons are strongly pinned at the interface and disorder is mainly due to the wide distribution of energy levels.

\begin{figure}
\begin{center}
\includegraphics[width=85mm,bb=3 1 241 128]{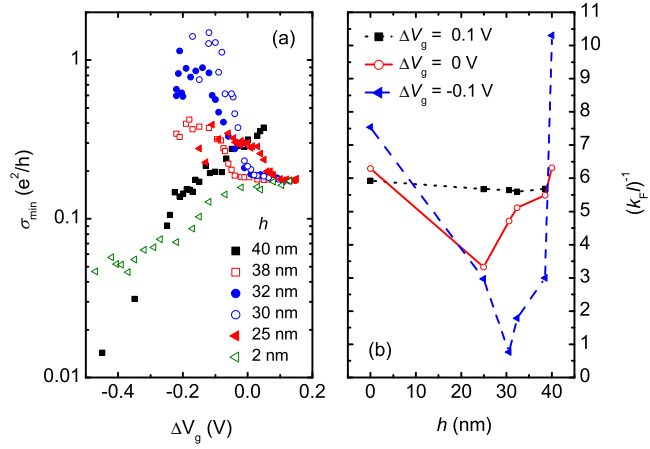}
\end{center}
\caption{\label{fig:figure2} a) Values of $\sigma_{\textup{min}}$ for different DCL positions. b) Variation of the disorder strength with the estimated DCL position in the oxide, at $\Delta V_{\textup{g}}\,=0.1$, 0 and -0.1\,V.}

\end{figure}

Previous experiments in sodium-doped MOSFETs reported an increase of disorder by either increasing the impurity charge concentration or by decreasing the electron density ($k_{\textup{F}}l\,\sim\,N_{\textup{ox}}^{-0.85}$ and $k_{\textup{F}}l\,\sim\,N_{\textup{s}}^{0.5}$).\cite{Pepper2} Our results follow the expected behavior when $N_{\textup{ox}}\geq N_{\textup{s}}$ i.e. when ionized or vacant sites are present but differ when there is an excess charge of electron at the interface, i.e. electron correlation gets stronger. The overall feature in figure 2b could be explained in terms of relative compensation between ionized impurity traps and electrons at the Si-SiO$_2$ interface. The disorder is minimum when all traps are neutral (one electron per ion on average) but increased if there is an relative excess of electron ($h\,>\,30$ nm, $N_{\textup{ox}}< N_{\textup{s}}$) or of ionized traps ($h\,<\,30$ nm, $N_{\textup{ox}}\geq N_{\textup{s}}$). The number of ionized traps is increased by decreasing $h$ whereas the number of electrons is controlled by the voltage applied to the top metallic gate. However the condition $N_{\textup{ox}}\sim N_{\textup{s}}$ corresponds to a local minimum of disorder as disorder seems to saturate to a lower value close to $k_{\textup{F}}l\,\sim\,1$ when $N_{\textup{ox}}\sim N_{\textup{s}}\ll 1$. Although for $N_{\textup{ox}}\sim N_{\textup{s}}$ there is an average of one localized electron per ion, the probability that a site gets ionized to form a doubly occupied state in a neighboring site is non-negligible. Such probability depends on the value of the ionization energy which in turns depends on $N_{\textup{ox}}$. By decreasing $N_{\textup{ox}}$, the ionization energy increases and single occupancy is favored and disorder minimized.

In conclusion, we have shown that the presence of sodium impurities in the oxide of a silicon MOSFET can be used to control and minimize the disorder at the Si-SiO$_2$ interface, by modifying their position in the oxide layer, whereas the electron-electron interaction is controlled by the electron concentration at the same interface. Disorder is competing with electron-electron interaction by modifying the amplitude of the potential fluctuations and the correlation length at the interface, leading ultimately to a crossover from a ES VRH to a Mott VRH in the electron density-disorder plane.

We would like to thank Drs T. Bouchet and F. Torregrosa from Ion Beam Services-France for the process in the device as well as funding from the U.S. ARDA through U.S. ARO grant number DAAD19-01-1-0552.

\end{document}